\documentclass[colorlinks=true, citecolor=blue, linkcolor=blue, urlcolor=blue]{aa}

\usepackage{graphicx}
\usepackage{comment}
\usepackage{txfonts}
\usepackage{orcidlink}
\usepackage{float}
\definecolor{purple}{rgb}{0.5,0,0.87}

\begin{document} 

\title{LIGHTS. Limitations of Far-Infrared Tracers for Galactic Cirrus Mitigation in the Era of Ultra-Deep Imaging}
\titlerunning{Limitations of Far-Infrared Tracers for Galactic Cirrus Mitigation}

\author{
Ouldouz Kaboud\orcidlink{0009-0007-7712-0683}\inst{1,4}
\and Ignacio Trujillo\orcidlink{0000-0001-8647-2874}\inst{2,3}
% ---------
\and Ignacio Ruiz Cejudo\orcidlink{0009-0003-6502-7714}\inst{2,3}
\and S.~Zahra Hosseini-ShahiSavandi\orcidlink{0000-0003-3449-2288}\inst{1}
\and Mireia Montes\orcidlink{0000-0001-7847-0393}\inst{4}
\and Sergio Guerra Arencibia\orcidlink{0009-0001-7407-2491}\inst{2,3}
% ------------------
\and Mauro D'Onofrio\orcidlink{0000-0001-8647-2874}\inst{1}
\and Dennis Zaritsky\orcidlink{0000-0002-5177-727X}\inst{5}
\and Javier Román\orcidlink{0000-0002-3849-3467}\inst{6}
% ------------------
\and Nushkia Chamba\orcidlink{0000-0002-1598-5995}\inst{7}
\and Giulia Golini\orcidlink{0009-0001-2377-272X}\inst{8}
% ------------------
\and Chen-Yu Chuang\orcidlink{0000-0003-2069-9413}\inst{5}
\and Minh Ngoc Le\orcidlink{0009-0003-0674-9813}\inst{2,3}
\and Carlos Marrero-de la Rosa\orcidlink{0009-0005-1728-8076}\inst{2,3}
\and Zahra Sharbaf\orcidlink{0009-0004-5054-5946}\inst{4}
% ------------------
\and Richard Donnerstein\orcidlink{0000-0001-7618-8212}\inst{5}
\and Sepideh Eskandarlou\orcidlink{0000-0002-6672-1199}\inst{9}
\and Raúl Infante-Sainz\orcidlink{0000-0002-6220-7133}\inst{9}
\and Garreth Martin\orcidlink{0000-0003-2939-8668}\inst{10}
\and Samane Raji\orcidlink{0000-0001-9000-5507}\inst{11,12}
\and Nafise Sedighi\orcidlink{0009-0001-9574-8585}\inst{2,3}
}

\institute{
Department of Physics and Astronomy, University of Padova, Vicolo Osservatorio 3, I-35122 Padova, Italy
\and Instituto de Astrofísica de Canarias, c/ Vía Láctea s/n, E-38205 La Laguna, Tenerife, Spain
\and Departamento de Astrofísica, Universidad de La Laguna, E-38206 La Laguna, Tenerife, Spain
\and Institute of Space Sciences (ICE, CSIC), Campus UAB, Carrer de Can Magrans, s/n, 08193 Barcelona, Spain 
\and Steward Observatory and Department of Astronomy, University of Arizona, 933 N. Cherry Ave., Tucson, AZ 85721, USA
\and Departamento de Física, Universidad de Córdoba, Campus de Rabanales, Edificio Albert Einstein, E-14071 Córdoba, Spain
\and NASA Ames Research Center, Space Science and Astrobiology Division M.S. 245-6, Moffett Field, CA 94035, USA
\and School of Physics, University of New South Wales, NSW 2052, Australia
\and Centro de Estudios de Física del Cosmos de Aragón (CEFCA), Plaza San Juan, 1, E-44001, Teruel, Spain
\and School of Physics and Astronomy, University of Nottingham, University Park, Nottingham NG7 2RD, UK
\and Departamento de Física Teórica, Atómica y Óptica, Universidad de Valladolid, 47011 Valladolid, Spain
\and Laboratory for Disruptive Interdisciplinary Science (LaDIS), Universidad de Valladolid, 47011 Valladolid, Spain
}

\date{Received 2026; accepted 2026}

%%%%%%%%%%%%%%%%%%%%%%%%%%%%%%%%%%%%%%%%%%%%%%%%%%%%%% Abstract %%%%%%%%%%%%%%%%%%%%%%%%%%%%%%%%%%%%%%%%%%%%%%%%%%%%%%%%%

\abstract
{Current and forthcoming deep optical imaging surveys are reaching surface brightness levels fainter than 30 mag/arcsec$^2$, where Galactic cirrus emission becomes a significant contaminant. This poses a major challenge for the study of extremely low-surface-brightness extragalactic structures. We investigate the relationship between far-infrared emission, as traced by the Herschel 250 $\mu$m band, and optical surface brightness in the Sloan \textit{g}- and \textit{r}-bands. We find that, for Galactic cirrus with $\mu_g\gtrsim$~27 mag/arcsec$^2$, a linear relation is consistent with the data, assuming optically thin dust, with optical cirrus emission amounting to $\sim$0.1\% of the far-infrared flux. Using Herschel far-infrared data to trace Galactic cirrus enables its identification over wide areas, but at the cost of significantly reducing spatial resolution. This limitation introduces source confusion and restricts the effective surface brightness depth that can be robustly probed in optical data. We show that, when cirrus emission is modeled using Herschel 250 $\mu$m data, the limiting surface brightness is $\mu_V\sim$~28 mag/arcsec$^2$, substantially shallower than the nominal depth of surveys such as LIGHTS or the 10-year coadd LSST ($\mu_V\sim$~30.5 mag/arcsec$^2$). These results highlight a fundamental limitation in the use of currently available far-infrared data for cirrus mitigation in ultra-deep optical imaging, and emphasize the need for higher-resolution cirrus tracers to fully exploit the depth of upcoming surveys.}

%%%%%%%%%%%%%%%%%%%%%%%%%%%%%%%%%%%%%%%%%%%%%%%%%%%%%%% keywords %%%%%%%%%%%%%%%%%%%%%%%%%%%%%%%%%%%%%%%%%%%%%%%%%%%%%%%%


\maketitle

%%%%%%%%%%%%%%%%%%%%%%%%%%%%%%%%%%%%%%%%%%%%%%%%%%%%% Introduction %%%%%%%%%%%%%%%%%%%%%%%%%%%%%%%%%%%%%%%%%%%%%%%%%%%%%%

\section{Introduction}

The study of low-surface-brightness (LSB) features in the Universe plays a key role in probing the hierarchical evolution of galaxies. The detection of faint objects \citep[e.g., ][]{1984AJ.....89..919S,1988ApJ...330..634I,1997AJ....114..635D,2005ApJ...631..208B,{2015ApJ...809L..21M,2016A&A...588A..89J,2018ApJ...863L...7M,2021A&A...656A..44R}}, stellar halos and tidal features around nearby galaxies \citep[e.g., ][]{2005ApJ...635..931B,2006MNRAS.365..747A,2008ApJ...689..936J, 2010AJ....140..962M,2015MNRAS.446..120D,2016ApJ...823..123T}, and intracluster light (ICL) in galaxy clusters \citep[e.g., ][]{1991ApJ...369...46U,2005ApJ...631L..41M,2005ApJ...618..195G,2010ApJ...720..569R,2014ApJ...781...24G,2014ApJ...794..137M,2022ApJ...940L..51M,2023A&A...679A.157R} collectively provide valuable evidence to test and refine the $\Lambda$CDM model \citep[e.g., ][]{1999ApJ...524L..19M, 1999ApJ...522...82K,2010MNRAS.406..744C}. 

However, LSB data are susceptible to systematic issues such as poor data reduction, incorrect sky subtraction, and contamination from instrumental scattered light caused by the  Point Spread Function (PSF) and internal reflections \citep{2019arXiv190909456M}. Fortunately, recent advances have reduced many of these barriers, thanks to two major developments. First, the improvements in instrument designs and observational techniques specifically optimized for LSB studies \citep[e.g., ][]{1996PASP..108..104Z,2012ApJS..200....4F,2014PASP..126...55A,2015MNRAS.446..120D,2016MNRAS.456.1359F,2018RNAAS...2..144R,2019MNRAS.490.1539R,2024MNRAS.529..197L}. Second, advancements in data analysis methods that preserve faint emission and reduce systematic errors \citep[e.g., ][]{2001ApJS..137..117G,2009PASP..121.1267S,2015ApJS..220....1A,2016ApJ...823..123T,2017ApJ...834...16M,2018ApJ...857..104G,2019A&A...621A.133B,2020MNRAS.491.5317I,2023MNRAS.520.2484K,2023ApJ...953....7L,2024MNRAS.528.4289W,2024eas..conf..455G,2025A&A...697A...6C}.

Despite these developments in instrumentation and data analysis that mitigate many systematic errors, the unavoidable presence of Galactic cirrus emission in the optical regime, caused by starlight scattered by dust grains \citep{2012ApJ...744..129B}, still poses a significant and distinct foreground contamination challenge in deep LSB imaging. Cirrus, which has been observed even at high Galactic latitudes by reflecting starlight in optical data \citep[][]{1937ApJ....85..213E,1941ApJ....93...70H,1976AJ.....81..954S,1979A&A....78..253M} has become unavoidable. The Galactic cirrus provides invaluable data for studies of the interstellar medium \citep{2016A&A...593A...4M}; however, it also acts as a foreground contamination that complicates the identification of faint extragalactic LSB sources. Due to its diffuse nature and sometimes similar morphology, cirrus can mimic ultra-diffuse galaxies and dwarf satellite galaxies, often occupying a significant fraction of the field of view and thereby creating confusion \citep{2021ApJS..257...60Z}. Moreover, when projected onto nearby galaxies, cirrus emission can be mistaken for collisional debris such as shells and tidal tails \citep{2010A&A...516A..83S,2020MNRAS.498.2138B}, and in clusters, it contaminates characterization of intracluster light \citep{2017ApJ...834...16M}. 

The problem of cirrus contamination is expected to be an even more serious issue in ongoing deep imaging surveys like the LBT Imaging of Galactic Halos and Tidal Structures (LIGHTS, \citet{2021A&A...654A..40T,2024AJ....168...69Z}), the Stellar Stream Legacy Survey \citep{2023A&A...671A.141M}, the Legacy Survey of Space and Time (LSST) 
\citep[see e.g.][]{2019ApJ...873..111I,2022MNRAS.513.1459M,2024MNRAS.528.4289W} and the Euclid Wide and Deep Surveys \citep[]{2022A&A...657A..92E,2022A&A...662A.112E,2025A&A...697A...1E}. For this reason, developing a methodology to identify and, if possible, remove these structures from the images is more urgent than ever.

Galactic cirrus has its peak emission in the far-infrared \citep[FIR; ][]{1984ApJ...278L..19L,2025arXiv250705217L} due to its low temperature \citep{2010ApJ...713..959V}. The FIR radiation has its origin primarily in the thermal emission from dust grains in equilibrium with the radiation field. Many observations have shown that the visible morphology of optical cirrus closely matches its FIR counterparts \citep{2025arXiv250705217L}, and that their emission intensities are well correlated \citep[e.g., ][]{2013ApJ...767...80I,2020A&A...644A..42R,2023MNRAS.524.2797M}. Based on this observed correspondence, optical diffuse emission associated with Galactic cirrus is generally expected to have a FIR counterpart, whereas optical diffuse emission without an infrared counterpart is less likely to originate from Galactic cirrus. This, for example, has been used to separate cirrus from ICL features \citep{2017ApJ...834...16M}.

While most previous works have focused on distinguishing Galactic cirrus from other astrophysical sources, comparatively fewer have addressed its actual removal from deep imaging data. Cirrus removal can significantly enhance the detection and analysis of faint extragalactic LSB features by reducing foreground contamination and background confusion  \citep[e.g., examples based on IRAS, Herschel and Planck data; ][]{1993AJ....106.1664A,2017ApJ...834...16M,2025ApJ...979..175L}. However, such removal methods pose challenges due to the need for accurate morphological decomposition, especially when cirrus structures overlap spatially and spectrally with galactic light. Moreover, even after removal, the data may become noisier, a trade-off that must be accepted to ensure the usability of deep observations at highly LSB levels.

This study examines the use of Herschel 250 $\mu$m data to mitigate Galactic cirrus contamination from ultra-deep imaging data. At this wavelength, the Herschel/SPIRE map has a pixel size of 6\arcsec, and its angular resolution, determined by the SPIRE beam, has a Full Width at Half Maximum (FWHM) of around 18\arcsec. Although this plate scale and spatial resolution is not as good as that of optical data, it is excellent for FIR facilities. Building on the approach in \citet{2017ApJ...834...16M}, we probe whether a linear relationship between the emission of Galactic cirrus in the FIR and the optical exists. We also explore how to accurately estimate the proportionality factor between the two datasets if the linear relation is valid. Other approaches, such as using the optical color of the cirrus \citep[see e.g.][]{2020A&A...644A..42R} and/or identifying the filamentary structure of the Galactic cirrus \citep[i.e.][]{2025ApJ...979..175L}, preserve the original spatial resolution of the optical image. However, when large galaxies and other linear structures, such as streams, are present in the images, these approaches can lead to severe confusion. This is because the appearance and/or colors of the cirrus and streams can be very difficult to distinguish when they are in the same line of sight. For this reason, we have chosen to sacrifice some spatial resolution in order to achieve a cleaner Galactic cirrus subtraction in this work.

The paper is organized as follows: Section~\ref{sec:data} introduces the datasets used in this study; Section~\ref{sec:meta} describes the methodology and cirrus removal procedure; Section~\ref{sec:results} presents the main results and Section~\ref{sec:con} provides a summary and discussion of our conclusions.
Unless otherwise stated, the magnitude system employed in this paper is AB \citep{1983ApJ...266..713O}. 

%%%%%%%%%%%%%%%%%%%%%%%%%%%%%%%%%%%%%%%%%%%%%%%%%%%%  DATASET %%%%%%%%%%%%%%%%%%%%%%%%%%%%%%%%%%%%%%%%%%%%%%%%%%%%%%%%%%%%%%

\section{Data}
\label{sec:data}

\subsection{NGC 3953}

\begin{figure*}
    \centering
    \includegraphics[width=0.9\textwidth]{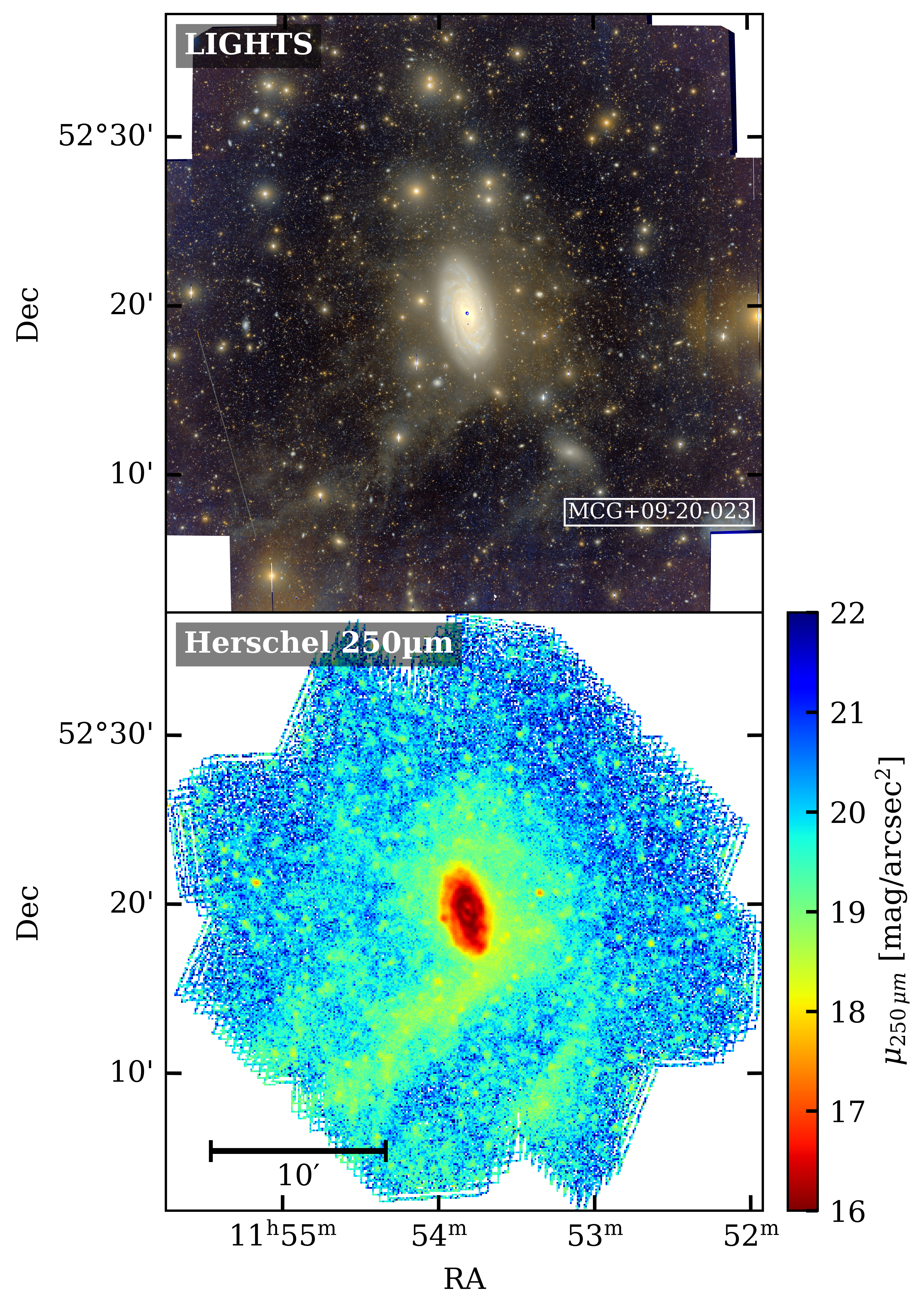}
    \caption{Upper panel: Color composite image of the NGC 3953 field, created using \textit{g}- and \textit{r}-band LIGHTS data. The diffuse structure around the galaxy is caused by Galactic cirrus. Bottom panel: A Herschel $250~\mu\mathrm{m}$ view of NGC 3953 and its surrounding area. The different colors in the image correspond to different surface-brightness intensities.}
    \label{fig:RGB}
\end{figure*}

To study the removal of Galactic cirrus contamination in ultra-deep optical imaging, we examined galaxy NGC 3953 and its surrounding region. Located at 18.8~$\pm$~1.0 Mpc \citep{2014ApJ...782...98B}, NGC 3953 is a SBbc  Milky Way–like galaxy with a maximum rotational velocity of 223~$\pm$~5 km/s \citep{2001A&A...370..765V}. 
We selected this galaxy because ultra-deep optical imaging data show that the area is severely contaminated by Galactic cirrus (see Fig.~\ref{fig:RGB}). This galaxy is particularly interesting because the location of the Galactic dust resembles merging activity. Therefore, it is worth exploring whether removing Galactic cirrus would reject this scenario. While the galaxy is particularly affected by Galactic cirrus, the region is not especially obscured, according to Galactic dust maps. In fact, based on the NASA/IPAC Extragalactic Database (NED), this region has a Galactic extinction \citep[A$_g$~=~0.099 mag, A$_r$~=~0.068 mag; ][]{1998ApJ...500..525S,2011ApJ...737..103S}. This field, compared with the rest of the LIGHTS sample (0.035$<$$A_g$$<$0.127 and 0.024$<$$A_r$$<$0.088), is the third most extinguished. This result shows that even at high Galactic latitudes ($b = 62.59\degr$; the LIGHTS sample's average Galactic latitude is $\sim70\degr$), the effect of Galactic cirrus is not negligible.

\subsection{Herschel far-infrared (FIR) data}

Following the approach and motivation of \citet{2017ApJ...834...16M}, we used Herschel data \footnote{\href{https://archives.esac.esa.int/hsa/whsa}{Herschel Science Archive }} to model the Galactic cirrus contribution in this field. NGC 3953 was observed by the Spectral and Photometric Imaging Receiver (SPIRE) instrument on the Herschel Space Observatory \citep{2010A&A...518L...3G} as part of the Herschel Reference Survey \citep{2010PASP..122..261B}. SPIRE produced images of this galaxy in three bands: $250~\mu\mathrm{m}$, $350~\mu\mathrm{m}$ and $500~\mu\mathrm{m}$, with pixel scales of 6.0, 10.0, and 14.0 arcsec/pixel, respectively. The angular (FWHM) resolution of each FIR filter is 18.1, 25.2, and 36.9 arcsec. Given these numbers, we compared our optical data with the $250~\mu\mathrm{m}$ band, which has higher angular resolution. This comparison is presented in Fig.~\ref{fig:RGB}.

Herschel images are calibrated and expressed in units of MJy/sr. For our analysis, we use the Level 2.0, zero-level-corrected, extended-source-calibrated SPIRE $250~\mu\mathrm{m}$ map (extdPSW) from the Herschel Science Archive, produced with the standard Herschel processing pipeline  \citep[SPG v14.1.0; ][]{Valtchanov2018}. Our goal is to create models of Galactic cirrus in the optical regime based on the spatial location of FIR emissions. The spatial resolution of the SPIRE images is limited by the beam size. The standard deviation of the background pixels in the  $250~\mu\mathrm{m}$ image is 0.63 MJy/sr (equivalent to $\sim$20.9 mag/arcsec$^2$). This limit is largely driven by the extragalactic confusion noise at this FIR wavelength due to unresolved sources within the SPIRE beam \citep[i.e. $\sigma_{\text{conf}}$~=~5.8~$\pm$~0.3 mJy/beam or, in the units given here, $\approx 0.6\,\text{MJy/sr}$;][]{2010A&A...518L...5N}.

\subsection{LIGHTS optical data}
\label{sec:LIGHTS}

NGC 3953 is part of the LIGHTS \citep{2021A&A...654A..40T,2024AJ....168...69Z} sample. LIGHTS is an ongoing ultra-deep imaging survey of nearby (D$\sim$15 Mpc) spiral galaxies with stellar masses slightly lower than the Milky Way. The LIGHTS observing and reduction strategy is designed to preserve as much as possible of the LSB features in the images. At the time of writing, the LIGHTS survey comprises a total of 28 nearby galaxies observed in the \textit{g}- and \textit{r}-bands simultaneously with the Large Binocular Cameras  \citep[LBC; ][]{2008A&A...482..349G} mounted on the Large Binocular Telescope (LBT) in Arizona. The LBC system consists of two cameras, Blue and Red, each composed of four Charge-Coupled Devices (CCDs) with a pixel scale of 0.224~arcsec/pixel. In the particular case of NGC 3953, the amount of time on-source allocated was 1.5h (of which 1.32h corresponds to the central region of the mosaic with more than 70\% of the exposure). This amount of time results in the surface brightness limits of 31.2 and 30.7 mag/arcsec$^2$ in the \textit{g}- and \textit{r}-bands, respectively ($3\sigma$ in areas equivalent to $10\arcsec \times 10\arcsec$). The LIGHTS optical images are originally expressed in nanomaggies, i.e., a zero point of 22.5 in the AB system, while the Herschel maps are in MJy/sr. We converted the nanomaggies units to MJy/sr for the LIGHTS \textit{g}- and \textit{r}-band images. The two main galaxies in the field are NGC 3953 and its satellite, MCG+09-20-023, located in the bottom-right part of the image (see Fig.~\ref{fig:RGB}).

Detailed explanations of the LIGHTS data reduction strategy can be found in  \citet{2021A&A...654A..40T} and \citet{2024AJ....168...69Z}. However, we will expand on how the optical sky background is estimated in these complex images, as removing this contribution is essential to preserving the low surface brightness of the optical cirrus. The LIGHTS low surface brightness-friendly pipeline uses a non-aggressive strategy to remove the optical background. Each frame used to generate the final co-added image is carefully masked, and the average value of the non-masked pixels is subtracted from each frame. Later, these frames are combined, and the co-added image's improved signal-to-noise ratio facilitates the detection of new low surface brightness signals. A new mask is created using the co-added image, and the improved masking is applied to the individual frames to better characterize the sky background in each frame. This process is repeated several times until a convergent sky subtraction is reached. In cases where fields are affected by Galactic dust, like in the case of NGC 3953, the Herschel's FIR image is used in addition to a purely optical-generated mask to create extra masking in the optical data. This avoids over-subtracting extremely low surface brightness optical dust emission.

\section{Methodology}
\label{sec:meta}

Before analyzing the data, we take several preparatory steps to enable direct comparison between the optical and Herschel data. First, we remove the scattered light produced by bright stars from the optical data. Second, we identify the spatial location of the Galactic cirrus. Next, we calculate the scaling factor to apply to the FIR data to match the optical dust emission. Finally, we subtract the Galactic cirrus from the optical data.

\subsection{Scattered light field removal in the optical dataset}
\label{sec:stars-sub}

The stars to be removed from the \textit{g}- and \textit{r}-bands of the LIGHTS images were selected from the Gaia catalog using the following two criteria:
a) Having a Gaia G$_{BP}$-band magnitude brighter than 15.5 mag, and
b) Being located at a radial distance from the central galaxy of less than 800 arcsec\footnote{In addition, we removed the scattered light from the three bright stars located at the following coordinates: R.A. $\sim$11h 52m and Dec $\sim$+52d 20m.}. The filter choice is based on its similar wavelength coverage to that of the combined Sloan \textit{g}- and \textit{r}-band response. The radial distance cutoff corresponds to the radial region covered by the Herschel field of view. Due to the dithering strategy, all the LIGHTS pixels within that radial range were observed at least 42\% of the time.

Scattered light subtraction was carried out using an extended PSF model appropriate for LBT/LBC observations, where both the PSF construction and subtraction procedure are described \citep{2025OJAp....8E..73S}. The model accounts for scattered light at large angular distances. Each optical band (\textit{g} and \textit{r}) has its own PSF model. These models were used to remove the contribution of the light of the stars in each band. The PSF model normalization was determined for each star by matching the observed radial surface-brightness profile of the model with the observed stellar profile. The scaled PSF was then shifted to the Gaia astrometric position of each star and subtracted iteratively, from the brightest to the faintest stars, using the Gnuastro \textit{NoiseChisel} and \textit{Segment} tools \citep{2015ApJS..220....1A,2019arXiv190911230A}. This procedure was applied independently to both the \textit{g}- and \textit{r}-band images to ensure consistent removal of stellar contamination.

\begin{figure*}[h]
    \centering
    \includegraphics[width=\textwidth]{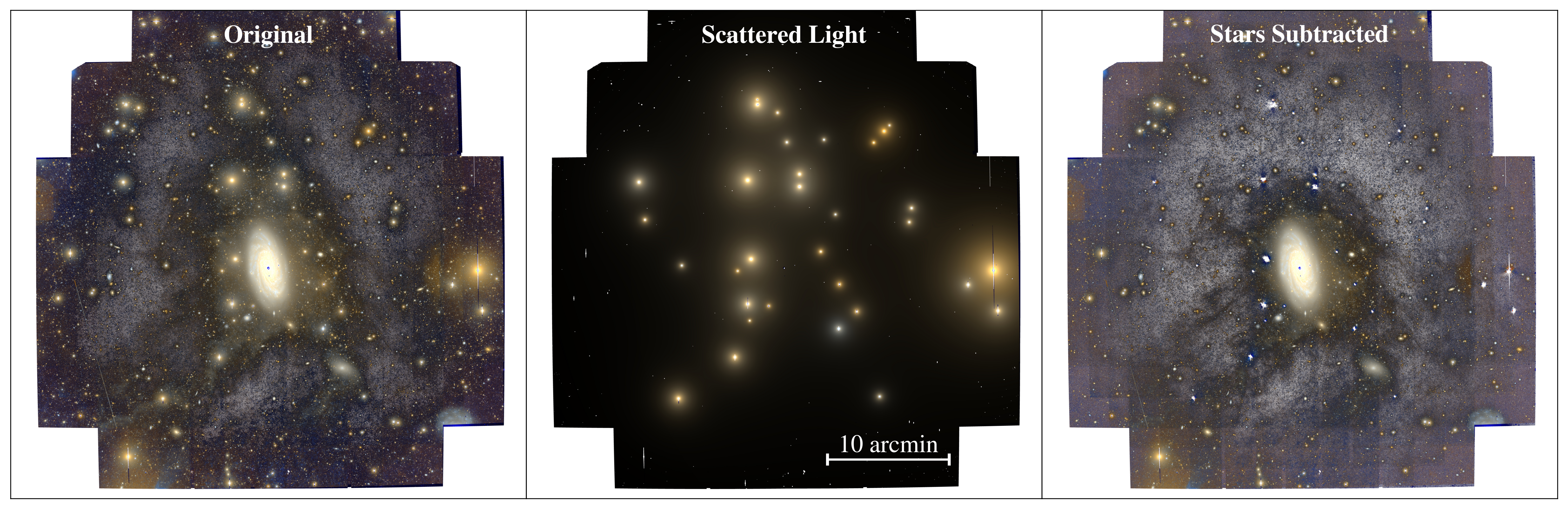}
    \caption{Modeling and subtracting the scattered light field produced by the brightest 28 stars (G$_{BP}$$<$15.5 mag) within an inner radial distance of 800 arcsec with respect to the position of NGC 3953. The three brightest stars in the western part of the image were also removed. The darker areas along the edges of the image are caused by the lower signal-to-noise ratio of the coadded mosaic in those regions.}
    \label{fig:Stars_subtract}
\end{figure*}

\begin{figure*}[h] 
    \centering
    \includegraphics[width=\textwidth]{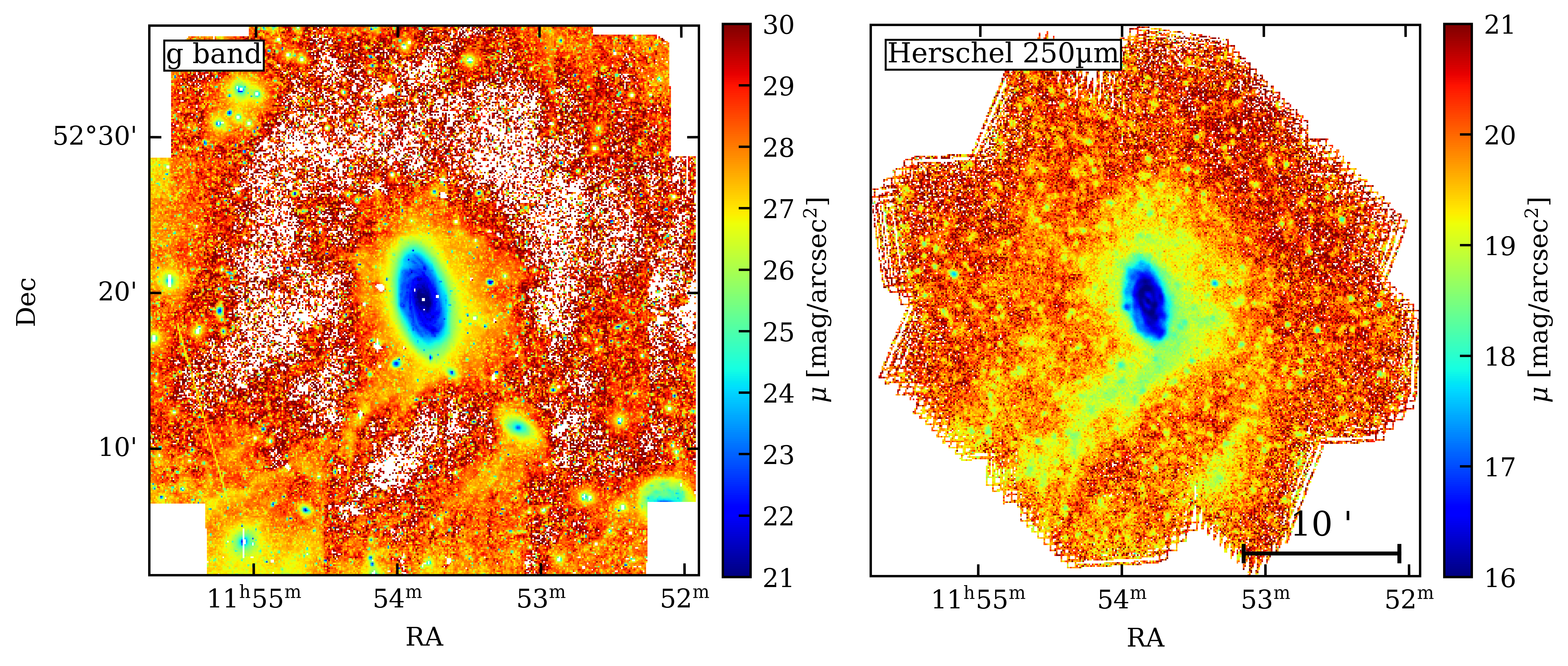} 
    \caption{A comparison of the optical scattered light-subtracted \textit{g}-band image (left) and the Herschel $250~\mu\mathrm{m}$ map (right). The diffuse emission features show clear spatial correlation in both filters, indicating that most of the faint optical background in this field originates from Galactic cirrus.}
    \label{fig:g_herschel}
\end{figure*}

A total of 28 stars were modeled and subtracted (S. Zahra Hosseini-ShahiSavandi, in preparation). Their spatial location is shown in Fig.~\ref{fig:Stars_subtract}. The same figure also shows the LIGHTS field before and after removing scattered light from these stars. At the location of the central galaxy, the contribution of scattered light to the image's surface brightness is significant: $\mu_g\sim$28.8 and $\mu_r\sim$27.9 mag/arcsec$^2$. These values are comparable to the surface brightness contribution of scattered light from bright stars in other deep LSB studies \citep[see e.g. ][]{2009PASP..121.1267S,2025OJAp....8E..73S}. This emphasizes the importance of removing the stellar scattered light component before analyzing the faintest surface brightness features of the images. After removing the scattered light, the spatial coincidence of Galactic cirrus in both the optical and infrared images is enhanced (see Fig.~\ref{fig:g_herschel}).

\subsection{Galactic cirrus isolation}
\label{sec:mask}

As previously mentioned, Galactic cirrus emission is prominent in the Herschel 250~$\mu$m band. Therefore, this data can be used to identify and isolate dust contamination in optical bands. However, the infrared data also shows light from sources unrelated to the Galactic cirrus. These sources include star-forming galaxies spanning a wide redshift interval, including high-redshift galaxies that appear as compact sources in the images and nearby galaxies that are more extended. Therefore, it is necessary to identify and mask those galaxies out of the analysis\footnote{The infrared emission from faint galaxies that cannot be resolved individually could contribute to the Cosmic Infrared Background (CIB) and be confused with Galactic cirrus emission. Some work has been conducted to separate the contributions of both components to the infrared \citep[see, for example, ][]{2016A&A...596A.109P}, thereby isolating the Galactic cirrus contribution. However, these maps have very poor angular resolution, so using them for our purposes is beyond the scope of this paper.}.

In the case of the central galaxy, NGC 3953, based on the optical extension, we mask all the pixels of the Herschel data within a radial distance of 350 arcsec from the center of the galaxy\footnote{The results reported in this paper do not show significant changes when the aperture of such masks is changed from 330 to 370 arcsec.}. At such radial distance, the surface brightness profile of the galaxy reaches a value of 27.5 mag/arcsec$^2$ (\textit{g}-band), which is well beyond the expected location of the edge surface brightness for Milky Way-like galaxies \citep[see Fig.~2 in ][]{2022A&A...667A..87C}. Next, we run MTO2\footnote{\href{https://github.com/m-faezi/MTO2}{GitHub repository of MTO2 }} \citep{10535192} and mask all the compact sources in the image (\textit{--move-factor=0} and \textit{--area-ratio=0.95}). We do some manual masking when necessary if some objects are not properly masked.

To properly compare with the Herschel data, the \textit{g}- and \textit{r}-bands from LIGHTS were rescaled to match the Herschel 250~$\mu$m plate scale. Before resampling the optical data, to account for the PSF of the Herschel data, the optical data were convolved with a Gaussian with an FWHM of 18.1 arcsec. To put everything into a common grid, we use the FIR image as a reference, and we apply \textit{astwarp} of Gnuastro.  Then, we masked the compact sources in the optical data using MTO2 and, when necessary, applied manual masking as well\footnote{We tested whether masking the compact sources in the optical images before resampling would significantly change the results of this work. We found that it did not. For this reason, we decided to use the same masking in both the optical and infrared images to ensure that the same type of contamination was removed.}. The optical and Herschel masks are combined to create a single final mask for both sets of data. The remaining pixels in the common area correspond to pixels with LSB and extended emission, i.e., Galactic cirrus and stellar halo. In addition, there are pixels dominated by noise in both datasets (i.e., background).

\begin{figure*}
    \centering
    \includegraphics[width=\textwidth]{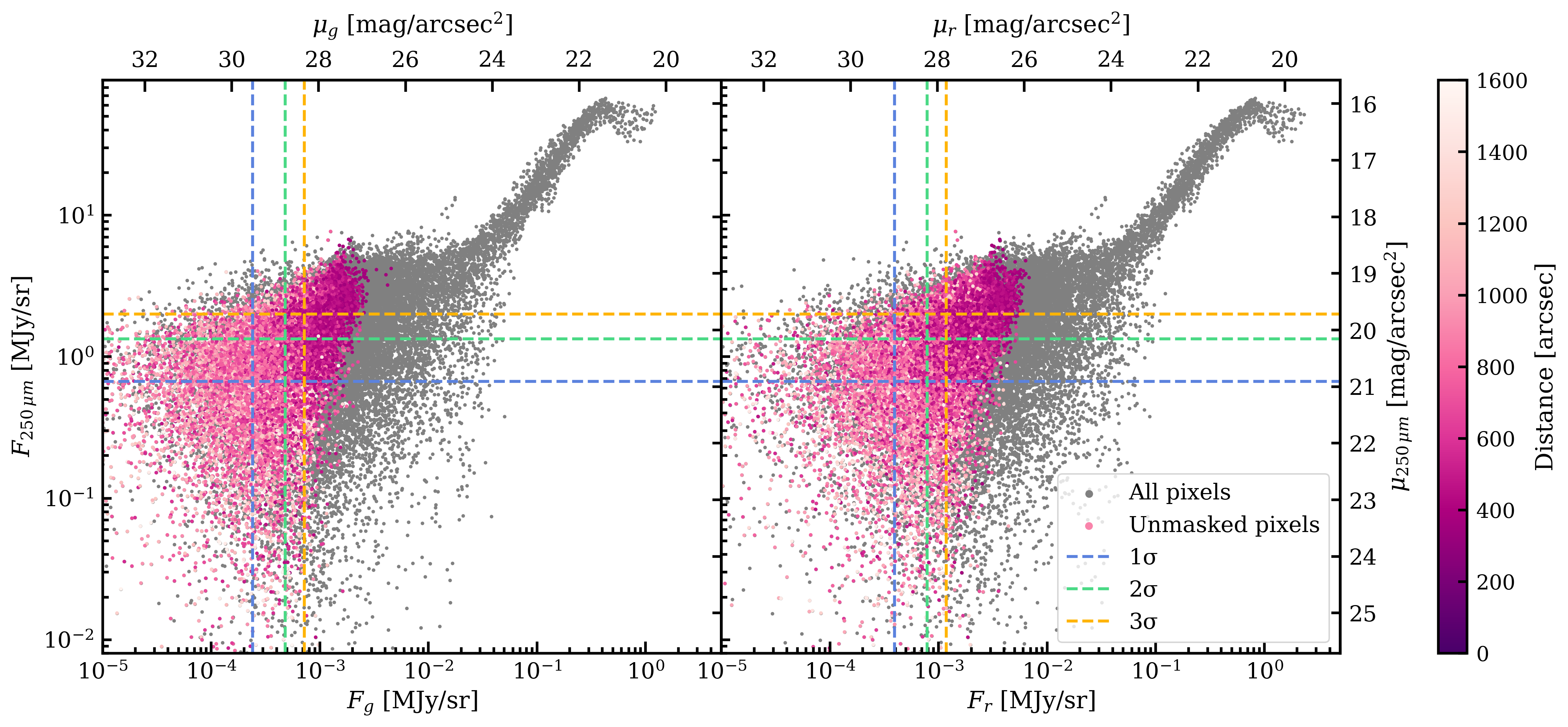}
    \caption{FIR ($250\,\mu\mathrm{m}$) vs optical (g, left; r, right) emission.  Each point represents emission from 6-arcsec pixels. The gray points represent all pixels, while the colored points (colored by their distance from the center of the NGC 3953 galaxy, i.e., the center of the image) correspond to pixels showing diffuse emission. The dashed lines indicate the 1, 2 and 3 standard deviations  levels above the respective background of the images.}
    \label{fig:Flux_Flux}
\end{figure*}

Fig.~\ref{fig:Flux_Flux} compares the flux of pixels surviving the aforementioned masking in 250~$\mu$m vs. optical band surface brightness plots. The position of all pixels in the image's common region is shown with gray points; colored pixels correspond to those that survive the masking process. The vertical and horizontal dashed lines indicate the equivalent surface brightness per pixel (i.e., $6\arcsec\times6\arcsec$) corresponding to one, two, and three standard deviations of the background. For the Herschel data, 1$\sigma$ is approximately 0.6 MJy/sr, or 20.9 mag/arcsec$^2$, likely reflecting contributions from instrumental noise and extragalactic and cirrus source confusion. Interestingly, in deep optical imaging, when using large pixel scales, as in this case, the surface brightness limits are also dictated by source confusion. In the optical, we find for 6 arcsec pixels, that 1$\sigma$ limit is approximately 2.4$\times$10$^{-4}$ MJy/sr, or 29.5  mag/arcsec$^2$ (\textit{g}-band), and approximately 3.1$\times$10$^{-4}$  MJy/sr, or 29.2 mag/arcsec$^2$ (\textit{r}-band).

The pixel emission distribution in Fig.~\ref{fig:Flux_Flux} exhibits several notable features. The most prominent one is the narrow branch that corresponds to the brightest pixels in both the optical and infrared images. Pixels with optical surface brightness ranging from around 20 to 24 mag/arcsec$^2$ are located on top of the galaxy NGC 3953. The region corresponding to optical surface brightness between 24 and 26.5 mag/arcsec$^2$ is more scattered, partly because this region is noisier, and partly because multiple sources contribute to it, such as background galaxies in the image. Additionally, this region includes the overlap of the stellar halo diffuse emission of the central galaxy and the Galactic cirrus.

As previously mentioned, the region in Fig.~\ref{fig:Flux_Flux} that is occupied by colored pixels corresponds to pixels that have not been masked. These pixels contain background pixels as well as diffuse emission far from the central galaxy. Consequently, they describe the correlation between FIR and optical emission for Galactic cirrus. Previous studies of high-latitude, optically thin dust clouds \citep[e.g. ][]{2008ApJ...679..497W,2013ApJ...767...80I} have shown a linear relation (in the log-log plane) between optical flux and infrared emission at 100~$\mu$m. This linear relation was also assumed by \citet{2017ApJ...834...16M} to remove Galactic cirrus contamination in optical data using FIR emission as a proxy. In addition, Fig.~3 of \citet{Zhang_2023} suggests that regions with 250~$\mu$m surface brightness below $\sim$10 MJy/sr lie approximately within the linear regime. The next subsection explores whether the almost triangular distribution observed in Fig.~\ref{fig:Flux_Flux} is compatible with a linear relationship between the two data sets.

\subsection{Scaling factor between FIR and optical data}
\label{sec:factor}

To test whether the FIR and optical emissions of the Galactic cirrus are proportionally related (i.e., $\alpha~$F$_{250\mu m}$~=~$F_{\mathrm{opt}}$), we created several models exploring different values of $\alpha$. To simulate the noise properties of our data, we consider Gaussian noise. This simple model can reproduce the observed relations between infrared and optical flux. 

\begin{figure*}
    \centering
    \includegraphics[width=\textwidth]{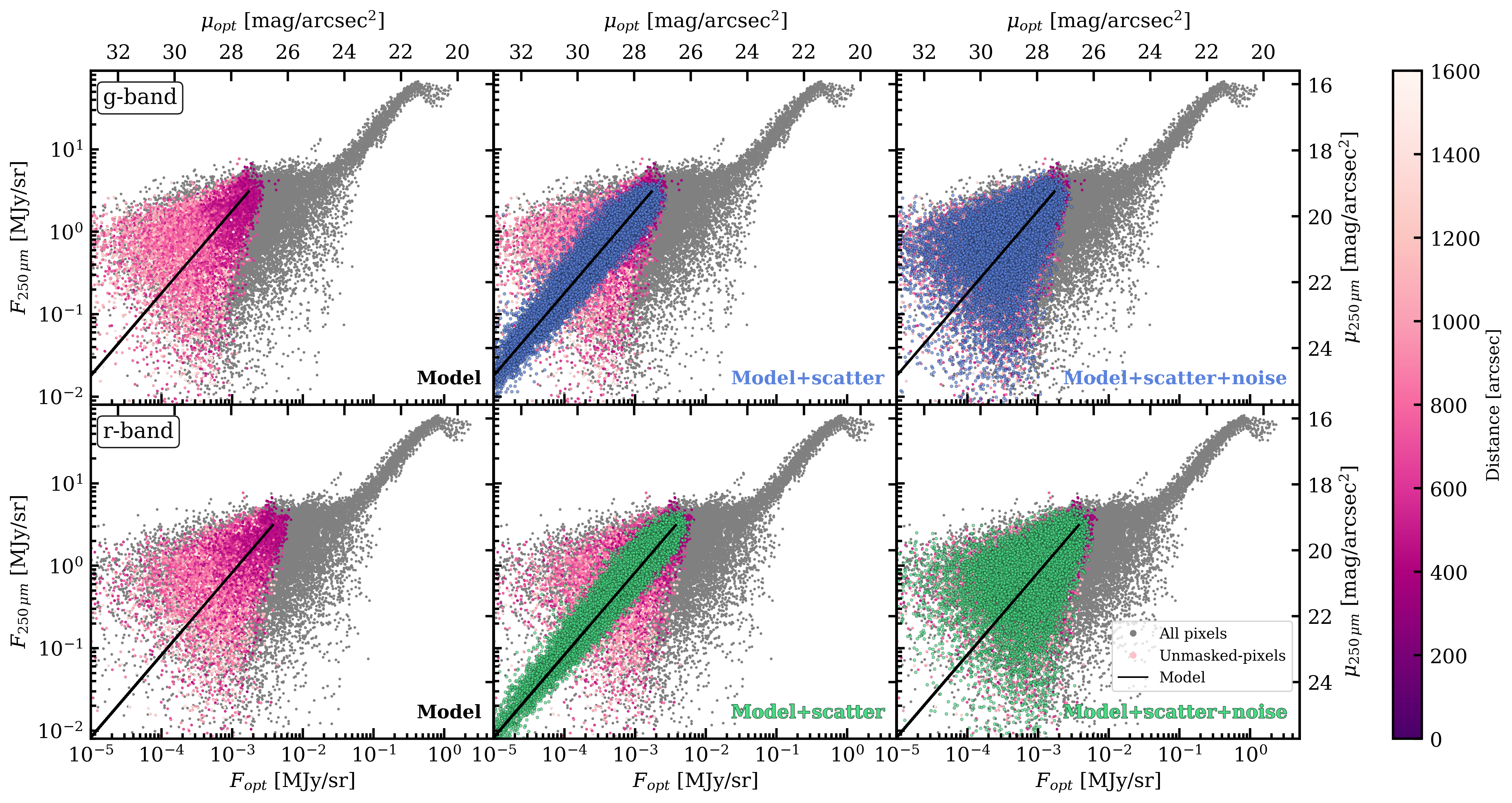}
    \caption{Similar to Fig.~\ref{fig:Flux_Flux} but this time showing the modeling approach we followed to reproduce the distribution of diffuse Galactic cirrus emission. The left panels show an example in which the infrared dust emission is proportional to the optical emission. The middle panels show the same relationship, but with intrinsic scatter proportional to the flux in each band. The right panels show the relationship after adding Gaussian noise with the properties of the infrared and optical backgrounds.}
    \label{fig:modeling}
\end{figure*}

The models were created as follows. First, we generated a couple of two-dimensional flux distributions: one for the infrared and one for each optical filter. These images have the same number of pixels as the number we want to simulate, i.e., the number of points in the colored triangular distributions shown in Fig.~\ref{fig:Flux_Flux}. The range of fluxes extends from zero to the peak flux observed in the triangular region. The two-dimensional gradient is modeled linearly. In practice, this means that there is a particular flux value at which the number of pixels is largest. This is done to model the fact that there are fewer pixels with low flux values, as well as fewer pixels with high flux values. We use the FIR distribution as a reference. Then we multiply that model by a given $\alpha$ value to obtain an optical model. The ratio of the two models is a constant by construction; therefore, plotting the flux of pixels in the optical and FIR maps produces a linear relationship (see left panel in Fig.~\ref{fig:modeling}).

Next, we add a Gaussian background to our models with the standard deviations measured in the 6 arcsec pixel images. After doing so, we find that the noise is insufficient to explain the scatter observed in the distribution of the colored points (see Fig.~\ref{fig:appendix1} in the Appendix~\ref{sec:appendix}). For this reason, we define a more realistic relation between the flux of the Galactic cirrus in the FIR and the optical.
As Fig.~\ref{fig:Flux_Flux} shows, the cirrus-contaminated pixels exhibit a scatter around the optical-FIR relation that is larger than that produced by the data's noise alone. This suggests that the relation has an intrinsic scatter, which may be due to variations in dust properties and local conditions, such as dust albedo, temperature, emissivity, and the illuminating radiation field. Consequently, we add an intrinsic scatter (proportional to the flux in each band) to our linear relationship in the models. This step is illustrated in the central panels of Fig.~\ref{fig:modeling}. Now, adding Gaussian noise to this more realistic relationship between the infrared and optical flux of Galactic cirrus reproduces the observed triangular distribution of colored points much more closely (see the right panel of Fig.~\ref{fig:modeling}).

Later, we quantitatively test whether we reproduce the observed data distribution with our simple model. To do that, we run a  two-dimensional Kolmogorov--Smirnov (KS2D) test\footnote{\href{https://github.com/Gabinou/2DKS?tab=readme-ov-file\#2-dimensional-kolmogorov-smirnov-test-for-goodness-of-fit}{GitHub repository of the 2DKS test}}. We conduct the KS2D test in log-log space to maximize the information used to determine the intrinsic scatter in the optical-infrared dust relationship. Doing the comparison in log-log space allows us to better describe the effect of noise at low flux levels and better estimate the intrinsic scatter, which we have assumed proportional to flux and therefore constant in log-log space. While this statistical test can not be used to identify the optimal $\alpha$ and intrinsic scatter that better reproduce the observed data distribution, it can be used to reject a combination of parameters that are not compatible between the models and the observations. The KS2D test returns a statistic D, such that smaller values of D indicate a larger similarity between the data and the models. The following combination of parameters provides the lowest D values. For the \textit{g}-band, an intrinsic scatter of 25.4\% in the \textit{g}-band and 19.3\% in 250~$\mu$m together with a ratio between the two fluxes of $\alpha_g$= 5.58($\pm$0.2)$\times$10$^{-4}$, gives D values as small as D=0.038. In the case of the \textit{r}-band, we find that an intrinsic scatter of 22.6\% in the \textit{r}-band and 18.6\% in 250~$\mu$m together with a ratio between the two fluxes of $\alpha_r$= 11.4($\pm$0.3)$\times$10$^{-4}$, gives D values as small as D=0.070. We find it remarkable that the intrinsic scatter is very similar in both the \textit{g}- and \textit{r}-bands. The ratio between the two proportionality factors, $\alpha_r/\alpha_g$=2.04$\pm$0.10, translates into a \textit{g}-\textit{r}=0.78$\pm$0.05. This color for the dust is compatible with those found by other authors \citep[see e.g.][]{2020A&A...644A..42R,2025ApJ...979..175L}. \citet{2017ApJ...834...16M} found (in the V-band) an $\alpha_V$ value 35($\pm$0.5)$\times$10$^{-4}$. This value is approximately 3 - 4 times larger than our estimates. The range of fluxes explored by \citet{2017ApJ...834...16M} is similar to ours in both the FIR and the optical.  We expect the difference to be related to  a number of reasons. First, the different methodologies and the trial-and-error approach \citet{2017ApJ...834...16M} followed. Second, the exact scaling factors are not expected to be universal from one field to another due to the different compositions of Galactic dust, grain size distribution, scattering phase function, inhomogeneity in the illumination, 3D scattering geometry, optical depth, and dust temperature/emissivity.

We finally run several tests in which we removed different background levels from our models to mimic potential oversubtraction of the real data. The models presented above assume that the background in the real data has been removed perfectly without any oversubtraction. By comparing models with oversubtraction to real data, we found that we are sensitive (using the KS2D) to alterations in the distribution of points in log-log space when we subtract optical fluxes larger than 10$^{-4}$ MJy/sr. Considering the typical optical sky background produced by our atmosphere of $\sim$0.3 MJy/sr (i.e., a surface brightness of $\sim$22 mag/arcsec$^2$), we could potentially detect oversubtraction in the real data (if any) at the $\sim$0.03\% level or higher.

\subsection{Galactic cirrus removal in the optical data}
\label{sec:removal}

Previous analyses have shown that the hypothesis of a linear relationship between dust emission in the optical and infrared is a satisfactory description for faint Galactic cirrus \citep[e.g. ][]{2017ApJ...834...16M,Zhang_2023}. In this section, we explore how to remove the dust contribution from optical images by scaling the Herschel 250~$\mu$m band to the expected optical flux. To accomplish this, we use the scaling factors determined in Section~\ref{sec:factor} and multiply the Herschel image accordingly. Unfortunately, there is a drawback to this approach. The removal of Galactic cirrus is limited by the angular resolution of the SPIRE data, set by the beam size (i.e., FWHM $\sim18''$ at $250\,\mu\mathrm{m}$ in this case). Therefore, the "dust-corrected" optical imaging has significantly poorer sampling after this correction. Additionally, as previously explained, the limiting surface brightness in optical data is also affected when using large pixels because source confusion is relevant.

\begin{figure*}[h]
    \centering
    \includegraphics[width=\textwidth]{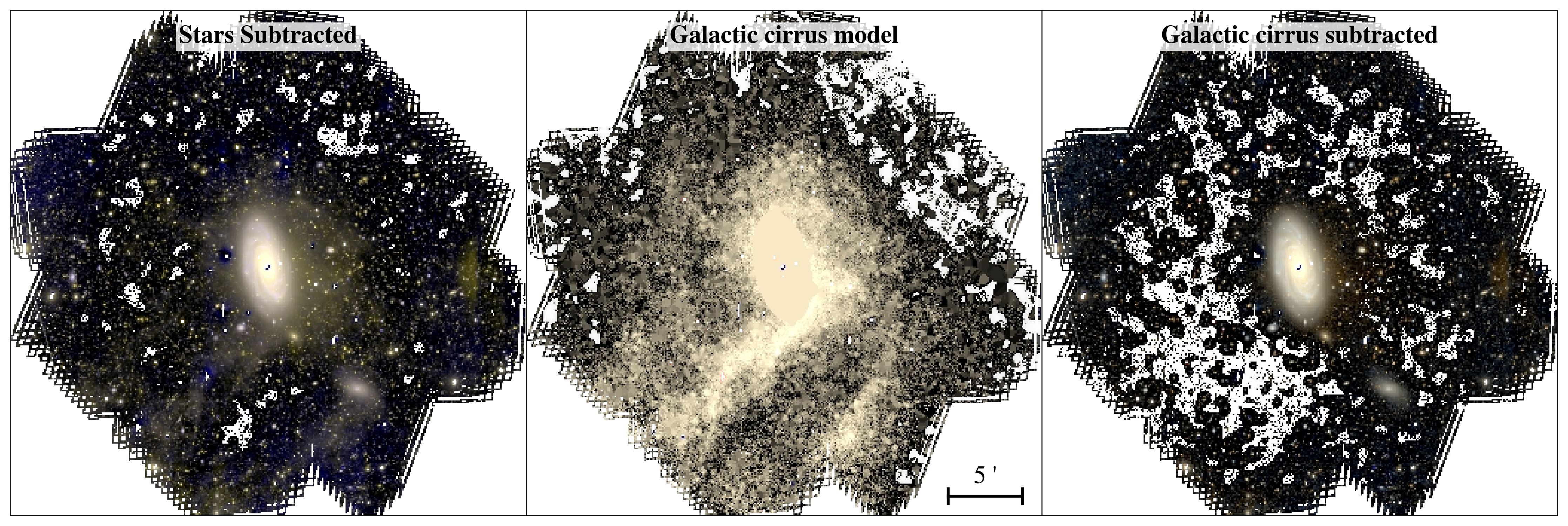}
    \caption{Modeling and subtracting Galactic cirrus in the LIGHTS NGC 3953 Field. The left panel shows the color optical image (at a 6$\arcsec$ pixel scale) after the scattered light field has been removed. The central panel shows the Herschel 250~$\mu$m image, in which the flux of the compact sources and the central galaxy has been adjusted to account for the expected dust contribution at those locations. The right panel shows the "dust-corrected" optical image after subtracting the central image from the one on the left. Background pixels are shown in white for better contrast.}
    \label{fig:result}
\end{figure*}

In addition to dust emission, the Herschel 250~$\mu$m image contains flux from high-redshift background objects and the central galaxy. Therefore, we need to fill in those locations with only the expected dust emission. For compact sources, such as background galaxies, we interpolate the flux of surrounding dust pixels in regions occupied by those objects. We use the Gnuastro \textit{astarithmetic} tool, \textit{interpolate-medianngb}, to perform this interpolation. This tool fills blank pixels occupied by high-z galaxies with the median value of the five nearest signal pixels. The values inside the blank region change depending on which pixels are closer. However, the central galaxy occupies a large part of the image, so interpolating as done before is not an adequate strategy. In this case, we measure the $250\,\mu\mathrm{m}$ surface brightness around the central galaxy using a 12\arcsec wide elliptical annulus with the same ellipticity and position angle as the central galaxy. We assume that the $250\,\mu\mathrm{m}$ surface brightness above the galaxy is similar to that in the surrounding region. Therefore, we model the dust brightness above the central galaxy with a fixed surface brightness. This step is illustrated in the middle panel of Fig.~\ref{fig:result}.

\begin{figure*}[h]
    \centering
    \includegraphics[width=\textwidth]{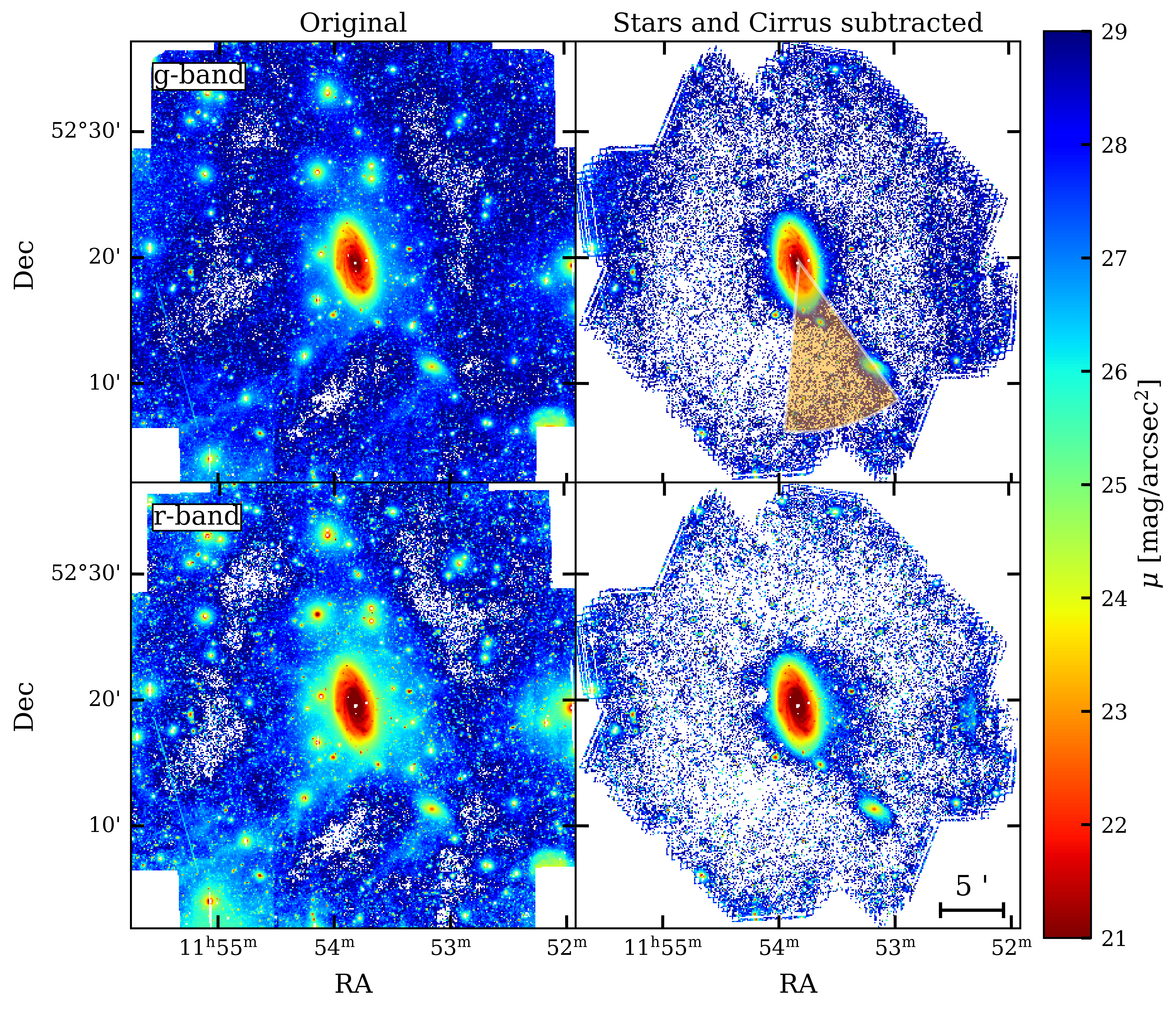}
    \caption{A comparison between the \textit{g}- and \textit{r}-band images before and after the removal of scattered light by bright stars plus the dust contribution. The edge of the central galaxy, NGC 3953, is now visible. The satellite galaxy,  MCG+09-20-023, located at the bottom right of the central object, now exhibits tidal features. The light excess at the right part of the image is a star out of the field of view. The upper right panel of the figure shows the position of the wedge-shaped region used to extract the radial profiles shown in Fig.~\ref{fig:SB}.}
    \label{fig:original-residual}
\end{figure*}

Once the infrared image is scaled to match the flux of the optical data and the pixels occupied by the compact and central galaxies are filled with their expected flux values, we subtract the infrared-scaled image from the optical image. The right panel of Fig.~\ref{fig:result} shows the LIGHTS NGC 3953 field after cirrus subtraction. Fig.~\ref{fig:original-residual} shows the "cleaned" \textit{g}- and \textit{r}-band images separately before and after the stars and dust removal. There are a number of results to highlight. First, the large envelope around the central NGC 3953 galaxy has disappeared. Second, the galaxy MCG+09-20-023, located at the bottom right of the central object, now shows tidal features. We will discuss these findings further in the next section.

\begin{figure*}[h]
    \centering
    \includegraphics[width=\textwidth]{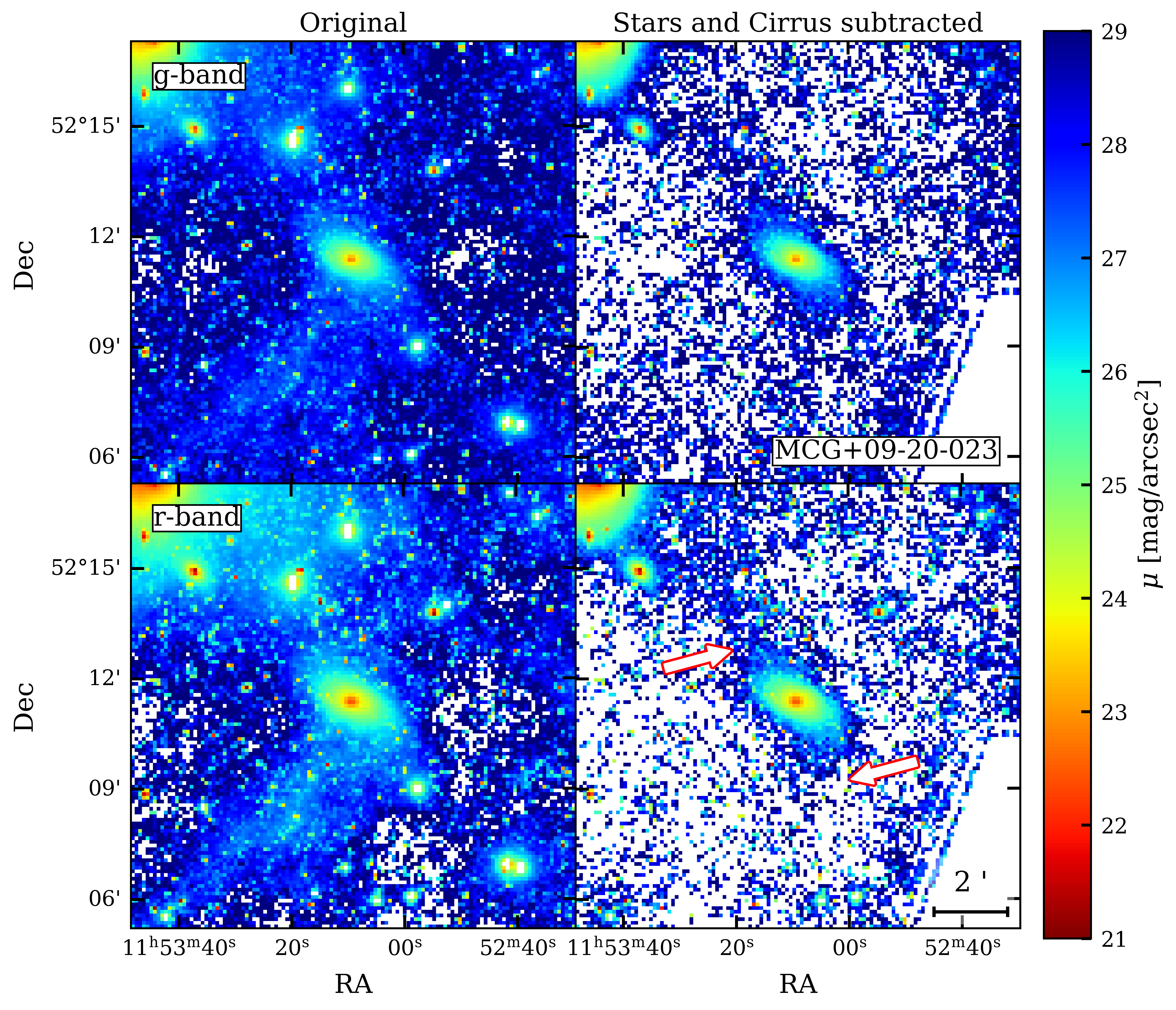}
    \caption{Similar to Fig.~\ref{fig:original-residual} but centered on MCG+09-20-023. 
         After removing the scattered light and the Galactic cirrus, a faint  S-shaped tidal structure around this galaxy becomes visible (indicated by arrows). This is strongly suggestive of an interaction with the central galaxy NGC 3953.
        }
    \label{fig:zoom}
\end{figure*}

%%%%%%%%%%%%%%%%%%%%%%%%%%%%%%%%%%%%%%%%%%%%%%%%%%%%  RESULTS %%%%%%%%%%%%%%%%%%%%%%%%%%%%%%%%%%%%%%%%%%%%%%%%%%%%%%%%%%%%%%
\section{Results}
\label{sec:results}

After removing scattered light of bright stars and Galactic cirrus, we quantitatively explore the effect of this unwanted light on the surface brightness profile of the central galaxy. We also probe the structural shape of the satellite galaxy MCG+09-20-023.

\subsection{The surface brightness and stellar mass surface density profiles of NGC 3953}
\label{sec:Radial-profile}

\begin{figure}[t]
    \centering
    \includegraphics[width=\columnwidth]{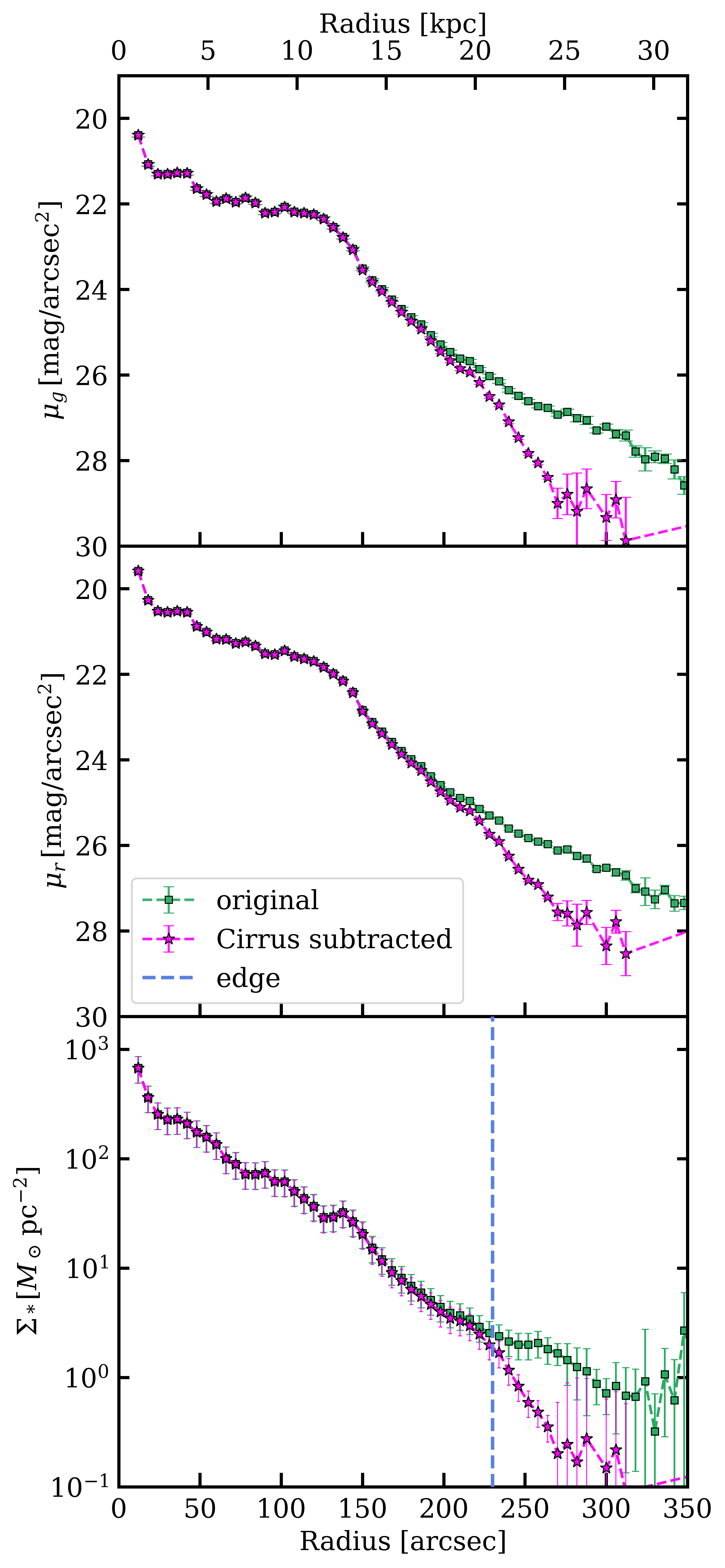}
    \caption{Radial surface brightness and stellar mass density profiles of NGC 3953 before (green) and after (magenta) Galactic cirrus subtraction. Profiles were derived using sigma-clipped mean averaging after masking background and foreground contaminant sources. Note how the edge of the galaxy at 230\arcsec\ (blue dashed line) only becomes visible after the cirrus removal.}
    \label{fig:SB}
\end{figure}

Fig.~\ref{fig:SB} shows the surface brightness and stellar surface mass density profiles of NGC 3953 before and after removing scattered light and Galactic cirrus. To extract the radial profiles, we used previous masks (excluding the central galaxy mask) to avoid contamination from foreground and background objects. We computed the surface brightness profiles in the \textit{g}- and \textit{r}-bands using the \textit{astscript-radial-profile} tool from Gnuastro \citep{2024RNAAS...8...22I}, adopting a photometric zero point of 22.5 mag. The center of the profile corresponds to the center of the galaxy (i.e. R.A.=178.45382, Dec=52.32677). The disk of NGC 3953 is well described with an axis ratio of b/a=0.50 and a position angle $102^\circ$ (counterclockwise from the X axis).

We extracted the radial profiles within a wedge-shaped region along the semi-major axis in the southern part of the galaxy (see Fig.~\ref{fig:original-residual}). This region is chosen because it is the less affected by the scattered light of the stars. To ensure robust averaging, we excluded outlier pixels in each radial bin using the sigma-clipped mean surface brightness option (\textit{--sigclip-mean-sb}). Fig.~\ref{fig:SB} displays how the resulting radial profiles are significantly affected by the Galactic cirrus correction. We corrected the profiles for galaxy inclination using the method described in \citet{trujillo2020physically}. After correcting the surface brightness profiles for the foreground Galactic extinction, we computed the stellar surface mass profile using the steps described in \citet{2008ApJ...683L.103B}. We derived the mass-to-light ratio using the prescription of \citet{roediger2015uncertainties} applied to the \textit{g}-band image and the BC03 stellar population models to obtain the coefficients.

Removing the Galactic cirrus reveals the edge of NGC 3953 visually at around 230\arcsec. The radial position, surface brightness, and stellar mass density at the edge of this galaxy are similar to those of other galaxies with comparable properties \citep[see e.g.][]{2022A&A...667A..87C,2025A&A...700A..91G}.

\subsection{Evidence of Tidal Interaction in MCG+09-20-023}
\label{sec:Tidal}
 
%----
In the original deep optical image, the satellite galaxy MCG+09-20-023 was heavily obscured by Galactic cirrus. This prevented us from seeing the LSB structures at the edge of the galaxy. After removing the scattered light and dust (see Fig.~\ref{fig:zoom}), the object revealed a distinctive S-shaped tidal feature. The shape of MCG+09-20-023 in its outskirts strongly suggests that it is being disrupted by NGC 3953. 

Such S-shaped tidal tails, bridges, or streams typically arise from tidal forces during close encounters, where the gravitational influence distorts the disk of the merging galaxy and produces extended features that trace the interaction history. Similar tidal structures have been documented in ultra-deep imaging studies, such as those of NGC1052-DF4 \citep{2020ApJ...904..114M,2022ApJ...935..160K,2024eas..conf..455G}.

%%%%%%%%%%%%%%%%%%%%%%%%%%%%%%%%%%%%%%%%%%%%%%%%%%%%  CONCLUSION %%%%%%%%%%%%%%%%%%%%%%%%%%%%%%%%%%%%%%%%%%%%%%%%%%%%%%%%%%%%%%
\section{Conclusions}
\label{sec:con}

Building on previous attempts to use FIR data as a direct tracer of Galactic cirrus location \citep[see e.g.][]{2017ApJ...834...16M,Zhang_2023}, we investigate whether a linear relationship between FIR and optical specific intensities accurately describes Galactic cirrus in our LIGHTS deep optical imaging. We confirm that this is the case for the range of Galactic cirrus surface brightness explored in this paper (i.e. $\mu_g>$27 mag/arcsec$^2$). This is expected, since optically thin cirrus is less affected by optical depth \citep{2025ApJ...979..175L}. To describe our data, we find that, in addition to assuming a linear relationship between optical and infrared surface brightnesses of the dust emission, it is necessary to assume an intrinsic scatter in the correlation (see Fig.~\ref{fig:modeling}). This scatter is estimated to be around 20\% of the Galactic cirrus flux. We interpret this value as resulting from slight inhomogeneities in the illumination received by the cirrus from the stellar population of our Galaxy, as well as from the range of sizes of the dust grains composing the interstellar medium, 3D scattering geometry, optical depth and/or dust temperature/emissivity.

We used the Herschel 250$\mu$m band to trace the location of the Galactic cirrus. This filter is a good compromise between moderate angular resolution (6$\arcsec$ pixel scale and $\sim$18$\arcsec$ of FWHM) and effectiveness in tracing the Galactic cirrus. In the surface brightness regime probed in this paper, the optical Galactic cirrus has around 0.1\% of the surface brightness of its FIR counterpart. However, using FIR data to trace the location of the Galactic cirrus sacrifices some of the spatial resolution of the optical dust-free image. This also limits the surface brightness features that can be explored in the optical. Increasing the pixel scale in deep optical imaging results in source confusion and a surface brightness "floor" that cannot be explored if we want to remove Galactic cirrus contamination from our images. In this paper, and for the particular case of using Herschel 250~$\mu$m to trace the Galactic dust, we estimate this limiting surface brightness to be around $\mu_V\sim$28 mag/arcsec$^2$. This value is approximately ten times worse than the expected value at the original pixel resolution for the LIGHTS survey (i.e., $\mu_V\sim$30.5 mag/arcsec$^2$; 3$\sigma$ in areas equivalent to $6\arcsec\times6\arcsec$). The main reason the depth of the optical, dust-free data has worsened is our inability to identify and mask individual background sources within the large pixel scale of the Herschel data. Some alternatives that avoid losing spatial resolution include those that use the color of optical Galactic cirrus and/or its filamentary morphology \citep[see e.g.][]{2020A&A...644A..42R,2025ApJ...979..175L}. These approaches are also imperfect when studying images with large, extended galaxies (i.e., several arcmin or larger) that potentially could also have stellar filamentary structures, such as stellar streams and shells.

For this particular study, we applied our technique to the LIGHTS NGC 3953 field. This field is severely affected by Galactic cirrus, and the external region of this object is overshadowed by the contribution of Galactic dust. After carefully removing both scattered light and cirrus, we can analyze the external region of NGC 3953 in detail and characterize its edge (where the star-forming region ends). Furthermore, we discovered that its satellite companion, MCG+09-20-023, is undergoing tidal disruption. 

In conclusion, although this methodology has some obvious drawbacks, it offers a practical and scalable solution for mitigating Galactic cirrus in ultra-deep optical surveys. It ensures more reliable photometric and morphological analyses, which will be increasingly important in upcoming wide-field, deep imaging campaigns. Overcoming the challenge of Galactic cirrus is essential for realizing the full potential of LSB science and achieving an unbiased view of the faint universe.
\clearpage
\begin{acknowledgements}

We would like to thank the referee for the detailed report, which helped improve the quality and clarity of the manuscript. The LBT is an international collaboration among institutions in the United States, Italy and Germany. LBT Corporation Members are: The University of Arizona on behalf of the Arizona Board of Regents; Istituto Nazionale di Astrofisica, Italy; LBT Beteiligungsgesellschaft, Germany, representing the Max-Planck Society, The Leibniz Institute for Astrophysics Potsdam, and Heidelberg University; The Ohio State University, representing OSU, University of Notre Dame, University of Minnesota and University of Virginia. Observations have benefited from the use of ALTA Center (alta.arcetri.inaf.it) forecasts performed with the Astro-Meso-Nh model. Initialization data of the ALTA automatic forecast system come from the General Circulation Model (HRES) of the European Centre for Medium Range Weather Forecasts. IT acknowledges support from the State Research Agency (AEI-MCINN) of the Spanish Ministry of Science and Innovation under the grant PID2022-140869NB-I00. This research also acknowledges support from the European Union through the following grants: "UNDARK" and "Excellence in Galaxies - Twinning the IAC" of the EU Horizon Europe Widening Actions programs (project numbers 101159929 and 101158446). Funding for this work/research was provided by the European Union (MSCA EDUCADO, GA 101119830). This work is part of grant CEX2025-001609-S, awarded to the Instituto de Astrofísica de Canarias under the Severo Ochoa Centre of Excellence program and funded by MICIU/AEI/10.13039/501100011033.
J.R. acknowledges financial support from the Plan Propio de Investigación 2025 submodalidad 2.3 of the University of Córdoba.
\end{acknowledgements}

% Use this for right citing
\bibliographystyle{aa}   % or apj, plainnat, etc.
\bibliography{myrefs}    % name of your .bib file (without .bib)
\clearpage
\onecolumn
\appendix

\section{Modeling the diffuse emission without intrinsic scatter}
\label{sec:appendix}
\begin{figure}[h]
    \centering
    \includegraphics[width=\linewidth]{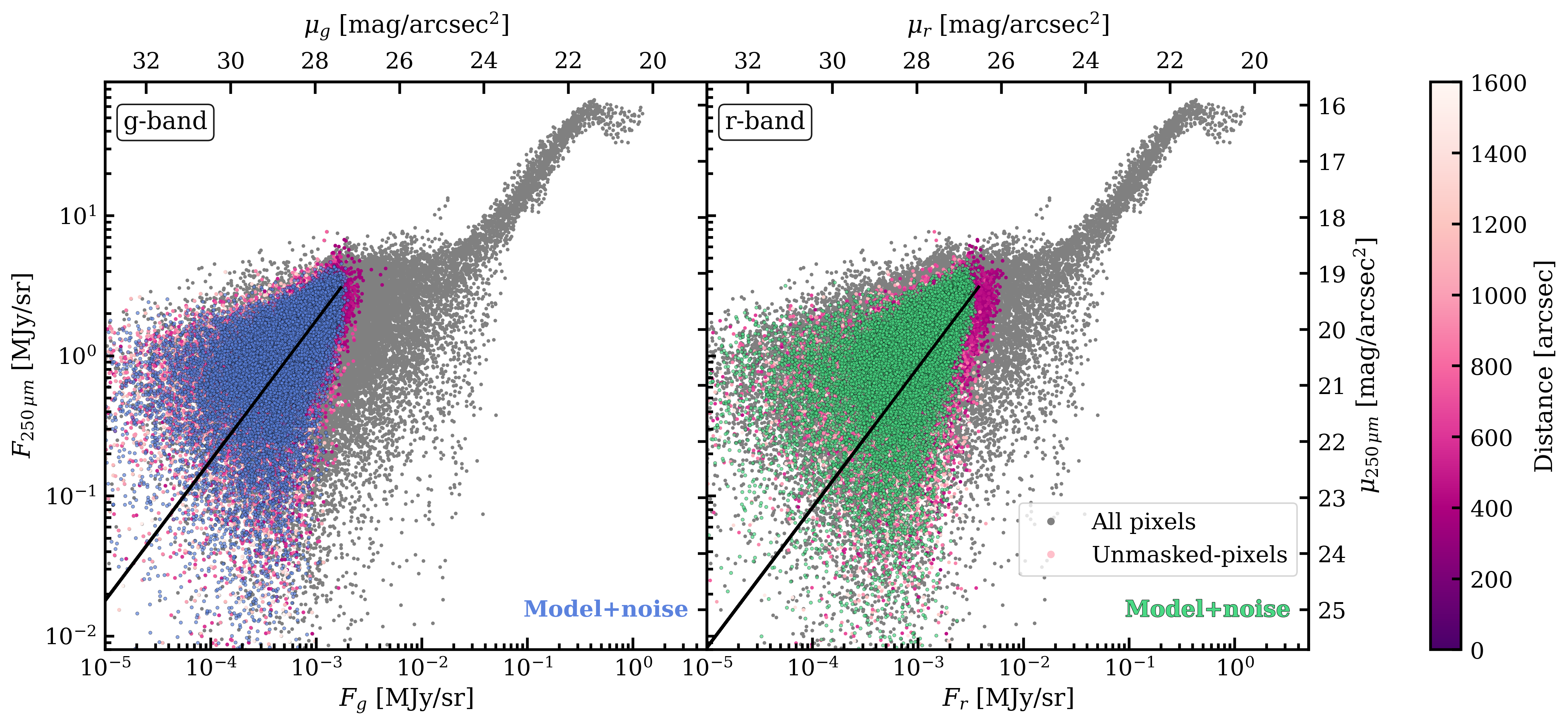}
    \caption{Modeling the relationship between dust flux in the optical and infrared. The black line represents a perfectly proportional relation between the optical and infrared fluxes. The effect of Gaussian noise, describing the noise properties of the images used in this work, is also shown. Pure Gaussian noise is not enough to explain the observed data distribution.}
    \label{fig:appendix1}
\end{figure}

Following the analysis in Section~\ref{sec:factor}, we present the results of adding only background noise to a perfect proportional relationship between dust flux in the optical and infrared in Fig.~\ref{fig:appendix1}. As shown in the figure, noise due to imaging quality alone does not fully explain the distribution of observed diffuse emission. In fact, the KS2D test yields larger D parameters (D=0.052 and D=0.075 for the \textit{g}- and \textit{r}- bands, respectively) than those found when an intrinsic scatter in the relation (of the order of 20\%) is included (see Section~\ref{sec:factor}).

% model+noise:
%D_r~0.075
%D_g~0.052

% model+noise+scatter
%D_r~0.070
%D_g~0.038

\end{document}